\documentclass[aps,prb,floatfix,superscriptaddress,twocolumn,nofitinbib,citeautoscript]{revtex4}
\usepackage{graphicx}
\usepackage{amsmath,bm}  
\usepackage{array}    
\usepackage{subcaption} 
\usepackage{amssymb}
\usepackage{booktabs}
\usepackage{braket}
\usepackage{amsfonts}
\usepackage{dsfont}
\usepackage{comment}
\usepackage{xcolor}
\usepackage{lipsum}
\usepackage{hyperref}
\usepackage{pifont}
\usepackage[justification=justified]{ragged2e}
\usepackage{etoolbox}
\makeatletter
\patchcmd{\@makecaption}
  {\centering}
  {\justifying}
  {}
  {}
\makeatother
\makeatletter
\long\def\@makecaption#1#2{%
  \vskip\abovecaptionskip
  \begingroup
    \justifying
    \sbox\@tempboxa{#1: #2}%
    #1: #2\par
  \endgroup
  \vskip\belowcaptionskip
}

\hypersetup{colorlinks=true,linkcolor=blue,anchorcolor=blue,citecolor=blue,filecolor=blue,urlcolor=blue,bookmarksnumbered=true,pdfview=FitB}

\begin{document}

\title{$p$-wave magnet driven field-free Josephson diode effect }

\author{Lovy Sharma}
\affiliation{Department of Physics, Indian Institute of Technology Delhi, Hauz Khas, New Delhi, India 110016}
\author{Bimal Ghimire}
\affiliation{Department of Physics, Indian Institute of Technology Hyderabad, Kandi, Sangareddy, Telengana, India 502285}
\author{Manisha Thakurathi}
\affiliation{Department of Physics, Indian Institute of Technology Hyderabad, Kandi, Sangareddy, Telengana, India 502285}
\date{\today}
\begin{abstract}
Recently, the superconducting diode effect (SDE), characterized by unequal critical currents in opposite directions, has been observed experimentally and predicted theoretically in models of bulk superconductors and Josephson junctions (JJs). In this work, we construct a Josephson junction using a recently discovered unconventional coplanar magnet, the $p$-wave magnet (PM), with proximity-induced superconductivity, and demonstrate the emergence of a Josephson diode effect (JDE). The barrier region is formed by another unconventional collinear magnet, namely an altermagnet (AM). We illustrate that apart from time-reversal and inversion symmetries, the mirror operation $M_{yz}$ emerges as the key symmetry constraint. Also, unlike earlier models that realize the JDE using unconventional magnets, this setup does not require Rashba spin–orbit coupling (SOC) or different superconductors across the junction. Moreover, we demonstrate that the realization of the JDE in this framework requires only minimal conditions while maintaining high performance. The effect remains robust across a broad parameter regime, and thus making the system particularly promising for applications in quantum circuits and computing technologies.

\end{abstract}
\maketitle

\section{Introduction}

Josephson junctions are among the most prominent, well-studied, and technologically important systems [\onlinecite{JOSEPHSON1962251,RevModPhys.51.101,RevModPhys.76.411,Acin_2018}]. Their importance lies not only in the macroscopic quantum nature, but also from their role as a cornerstone of superconducting electronics and quantum computing, for example, transmon qubit [\onlinecite{PhysRevA.76.042319,DiCarlo2009,Majer2007}], flux qubit [\onlinecite{RevModPhys.73.357}], charge qubit [\onlinecite{Nakamura1999}], and SQUID [\onlinecite{PhysRev.141.367}]. Recently another exotic phenomenon has been observed in the JJs, namely Josephson diode effect (JDE)  [\onlinecite{Ando2020,PhysRevLett.99.067004,PhysRevB.103.245302,PhysRevB.106.214524,doi:10.1126/sciadv.abo0309,mondal2025josephsondiodeeffectandreev,PhysRevB.109.174511,Soori_2025,Debnath_2025,sahoo2025fieldfreetransversejosephsondiode}], where the supercurrent flows in one direction and resistive current in the opposite direction due to the different de-pairing current (critical current). This phenomenon is similar to the non-reciprocity exhibited by a $p-n$ junction diode [\onlinecite{6773080}], where current experiences low resistance in the forward direction and high resistance in the reverse direction. JDE has potential applications in various modern technologies, such as superconducting logic circuits, quantum processors, direction selective quantum sensors and rectifiers [\onlinecite{inglaaynes2024highlyefficientsuperconductingdiodes,Golod2022,mohebi2025overviewjosephsonjunctionsbased}]. Non-reciprocal supercurrent has also been demonstrated in bulk superconductors. The first experimental observation was reported in a bulk $N_b/V/T_a$ super-lattice subjected to an external magnetic field [\onlinecite{Ando2020}]. Since then, considerable theoretical and experimental studies have verified the existence of the diode effect in both bulk superconductors and JJs [\onlinecite{PhysRevApplied.22.064017,PhysRevLett.129.267702,PhysRevLett.128.037001,PhysRevX.12.041013,Wu2022,PhysRevB.110.014518,PhysRevLett.130.266003}].

The necessary, though not sufficient, conditions for the JDE are the breaking of time-reversal symmetry (TRS) and inversion symmetry (IS) [\onlinecite{doi:10.1126/sciadv.abo0309,shaffer2025theoriessuperconductingdiodeeffects}]. In conventional JJs, TRS is typically broken either by applying an external magnetic field or by incorporating a ferromagnetic element into the junction. However, this approach poses challenges for practical applications, particularly in quantum computing, since stray magnetic fields arising from finite magnetization can disrupt device operation and increase noise. Recently, several theoretical and experimental studies have demonstrated the JDE without the use of external magnetic fields or ferromagnets; this phenomenon is called the field-free JDE [\onlinecite{PhysRevB.110.014518,Wu2022,Debnath_2025,hou2025fieldfreejosephsondiodetunable,ruthvik2025fieldfreediodeeffectsonedimensional}]. One promising route to break TRS without inducing finite magnetization, which we also employ in this work, involves the use of unconventional magnets (UMs) [\onlinecite{PhysRevB.110.014518,yqsg-xdg8,sahoo2025fieldfreetransversejosephsondiode}]. UMs are recently discovered magnetic phases that go beyond the traditional classification of magnets. Their classification is based on spin-group symmetry, in which the lattice and spin transformations are treated independently. For example, altermagnets (AM) break TRS, similar to a ferromagnet, yet exhibit zero net magnetization as in an antiferromagnet. In this phase, opposite spin sublattices are related by rotational symmetry in both real and momentum space, leading to alternating spin Fermi surfaces [\onlinecite{PhysRevX.12.040002,PhysRevX.12.040501,PhysRevX.12.031042,PhysRevX.12.011028,PhysRevX.15.021083,Ma2021}].
Along similar lines, another unconventional phase of coplanar magnetism, referred to as a $p$-wave unconventional magnet (PM), has also been proposed [\onlinecite{hellenes2024pwavemagnets}]. PMs possess zero net magnetization but exhibit spin splitting in momentum space with 
$p$-wave symmetry, thereby breaking IS. In PMs, opposite spin sublattices are connected by a combined time-reversal operation and a fractional lattice translation. Due to their anisotropic spin splitting, these systems display several exotic phenomena when interfaced with superconductors [\onlinecite{PhysRevB.111.035404,PhysRevB.111.165413,PhysRevB.111.174519,doi:10.7566/JPSJ.93.114703,Fukaya2025,PhysRevB.111.L220403,PhysRevLett.133.226002,heras2025interplaysuperconductivityaltermagnetismdisordered,pal2025topologicalsuperconductivitysuperconductingdiode,lu2026engineeringsubgapstatessuperconductors,fukaya2026crossedsurfaceflatbands,fukaya2025pwavesuperconductivityjosephsoncurrent,10.1088/1361-648X/ae430e}], such as spin-polarized transport [\onlinecite{PhysRevB.111.035404,10.1088/1361-648X/ae430e}], crossed Andreev states [\onlinecite{PhysRevB.111.165413}], topological superconductivity [\onlinecite{PhysRevB.111.174519,pal2025topologicalsuperconductivitysuperconductingdiode,fukaya2026crossedsurfaceflatbands}],  and $\phi_0$ JJs  [\onlinecite{PhysRevLett.133.226002}].

\begin{figure}[t]
  \begin{subfigure}{0.9\linewidth}
    \includegraphics[width=\linewidth]{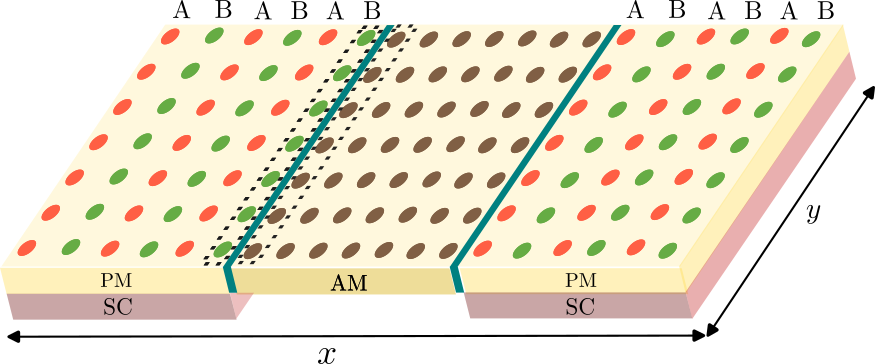}
    \caption{}
    \label{fig:cluster-a}
  \end{subfigure}
  \vspace{0.6em}

  \begin{subfigure}{0.49\linewidth}
    \centering
    \includegraphics[width=\linewidth]{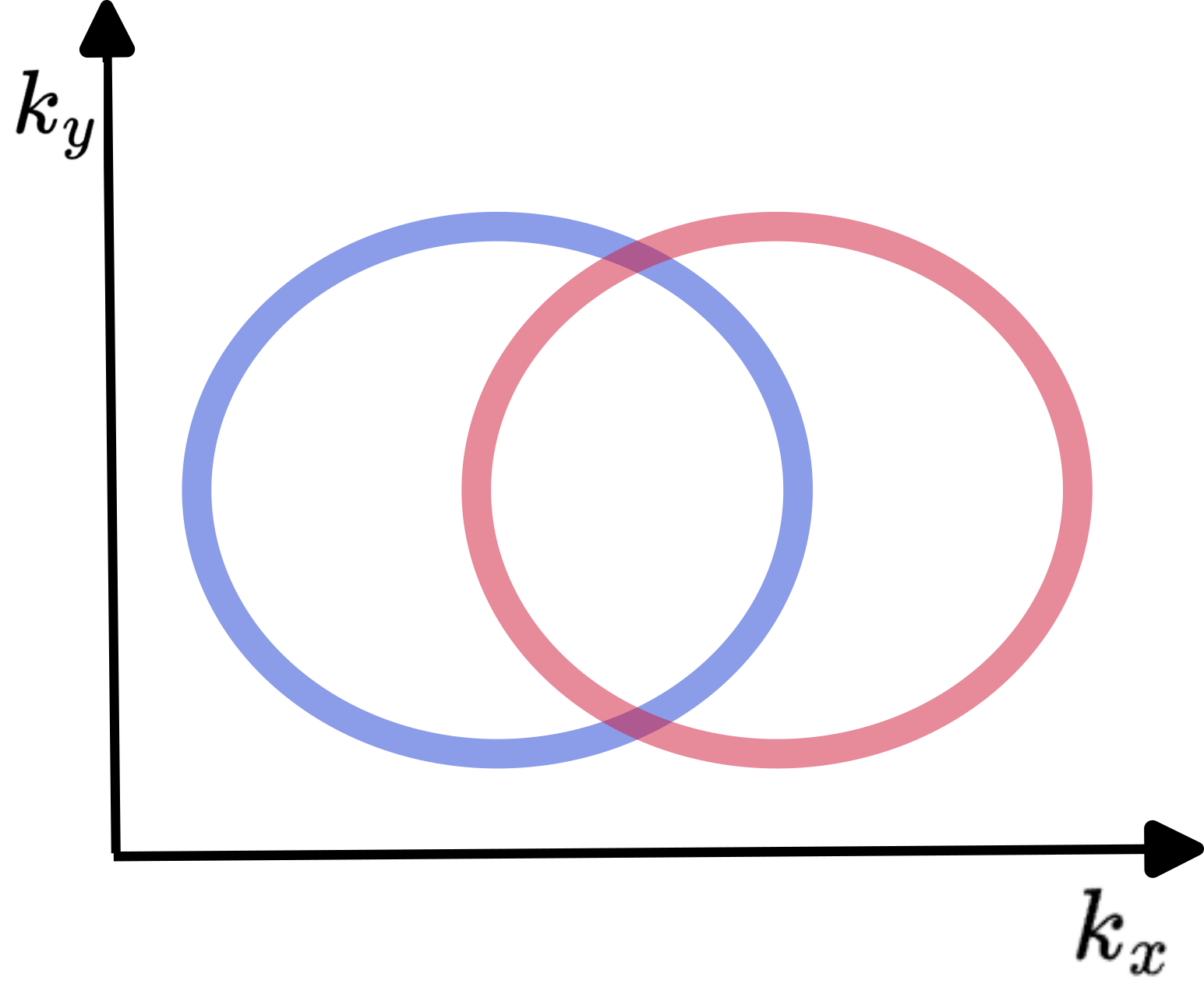}
    \caption{}
    \label{fig:cluster-b}
  \end{subfigure}\hfill
  \begin{subfigure}{0.49\linewidth}
    \centering
    \includegraphics[width=\linewidth]{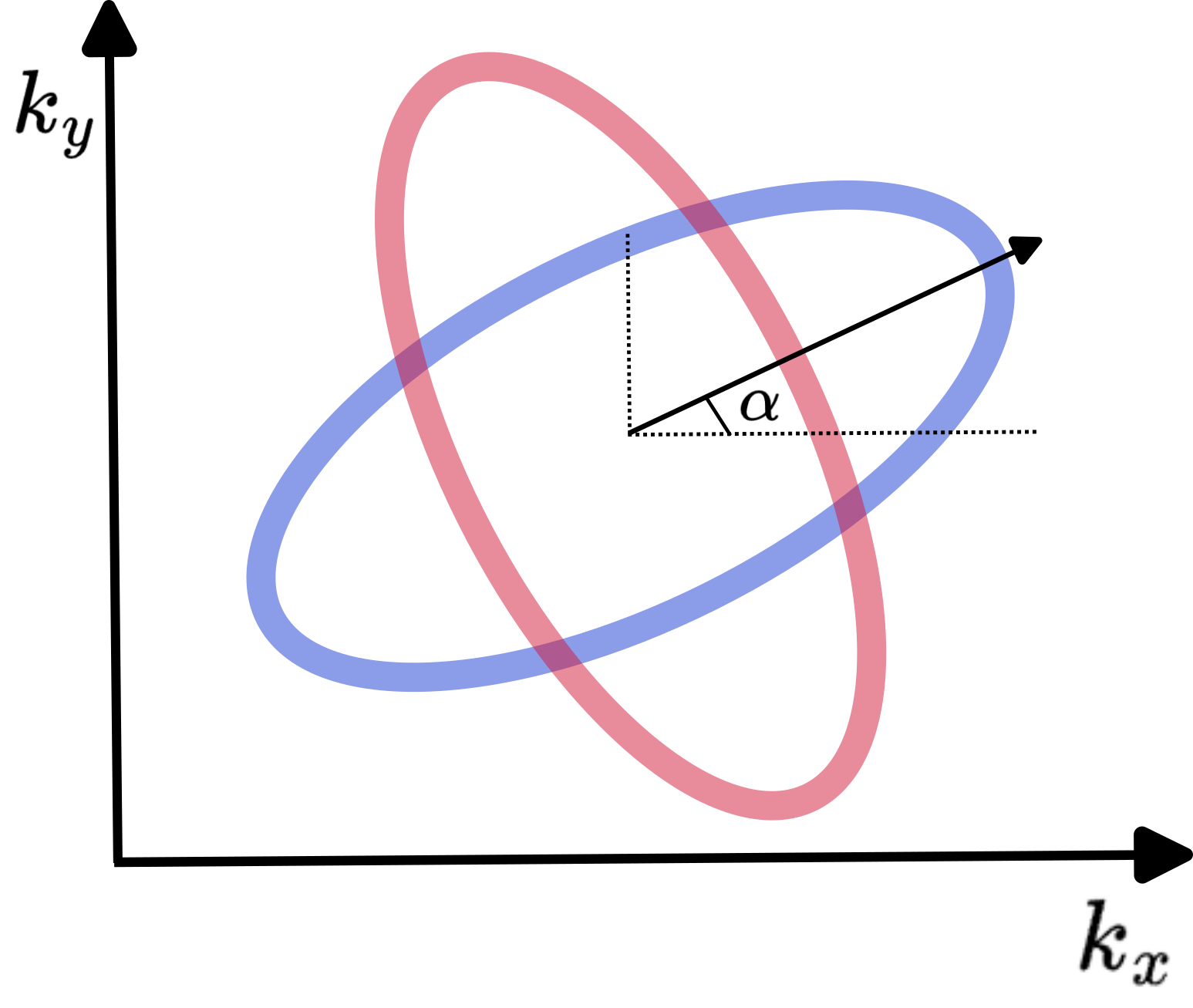}
    \caption{}
    \label{fig:cluster-c}
  \end{subfigure}

  \caption{(a) Schematic representation of a JJ consisting of the PMSC leads separated by an AM  barrier. The red and green lattice sites in the PMSC denote the `A' and `B' sublattices, respectively, while the black lattice sites belong to the AM region. Tunneling between PMSC and AM is assumed to be nearest neighbor only and hence, hopping is in AM lattice sites and their adjacent PMSC sublattice sites. (b) and (c) show the schematic Fermi surfaces of the PMSC and AM, respectively, where the red and blue contours correspond to spin-up and spin-down bands.}
  \label{fig1}
\end{figure}
  Motivated by recent progress and the emergence of novel phenomena in UMs, it is natural to explore heterojunctions involving AMs and PMs. In this work, we consider a JJ in which proximity-induced superconductivity in the PM serves as the SC leads, while the AM acts as the barrier.  Besides TRS and IS, mirror symmetry ($M_{yz}$) is found to be essential for the appearance of JDE. We also show that even with identical SCs on both sides of the junction, and in the absence of Rashba SOC and external magnetic fields, the system exhibits nonreciprocal current, in contrast to previous AM-based field-free Josephson diode proposals where Rashba SOC was essential [\onlinecite{yqsg-xdg8}]. We further demonstrate that introducing Rashba SOC in the AM allows one to tune both the magnitude and the polarity of the nonreciprocity in current. Moreover, a high efficiency of non-reciprocity is achieved over a wide range of exchange-field strengths and other system parameters. Thus, the planar Josephson junctions considered here exhibit the diode effect under fewer constraints than previous proposals based on unconventional magnets, to the best of our knowledge.
 
 The paper is organized as follows. In Sec. [\ref{sec1}], we
present the model of the planner JJ being used, the Hamiltonian of each element and algorithm used to calculate current. In Sec. [\ref{sec3}], numerically calculated results are shown for different parameter ranges of exchange field strength, Rashba SOC and crystallographic lobe angle. In Sec. [\ref{sec4}] we present the symmetry analysis of the result, explaining the necessary conditions for JDE. Then finally in Sec. [\ref{sec5}] we conclude our analysis.

\section{Model and Hamiltonian}
\label{sec1}
We consider a PMSC-AM-PMSC, 2D planner Josephson junction (JJ), where the left and right leads of PMSC are formed by coupling PM with a bulk SC and AM as a barrier (see Fig. \ref{fig1}(a)). The junction extends along the $x-$axis. The length of the left (right) SC lead is  $L^{L}_x=N^{L}_x a$ ($L^{R}_x=N^{R}_x a$) and is $L^{AM}_x=N_x^{AM}a$ for AM, where $a$ is the lattice spacing and $N_x^{L/R/AM}$ is the number of lattice sites along the $x-$axis. The width of the junction is $L_y=N_ya$ which is considered identical in all regions of the junction.  The complete Hamiltonian of the junction has the form $\hat{H}^{JJ}= \hat{H}^\text{PMSC}+ \hat{H}^{AM}+\hat{H}^{C_L} + \hat{H}^{C_R}$. Here, $\hat{H}^\text{PMSC}$, $\hat{H}^{AM}$ are the Hamiltonian for PMSC, AM respectively, whereas $\hat{H}^{C_L}$ ($\hat{H}^{C_R}$) is the coupling Hamiltonian between left (right) SC and barrier.
The form of $\hat{H}^{C_{L/R}}$ is described in Appendix [\ref{appA}] for the junction shown in Figs. \ref{fig1}(a). Next, we describe the form of $\hat{H}^\text{PMSC/AM}$ in detail as follows. 

We use a minimal model of PM to construct the Hamiltonian of PMSC [\onlinecite{hellenes2024pwavemagnets}]. The Bogoliubov-de Gennes (BdG) Hamiltonian for PMSC region in periodic boundary condition is given by, 

\begin{align}
    \hat{H}^\text{PMSC}=&\sum_k[\Psi^{\dagger}_{A,k},\Psi^{\dagger}_{B,k}]H^\text{PMSC}(k)[\Psi_{A,k},\Psi_{B,k}]^{T},
    \label{eq1}
\end{align}
 where  $\Psi_{X,k}=[\psi_{X,\uparrow,k},\psi_{X,\downarrow,k},\psi^{\dagger}_{X,\uparrow,-k},\psi^{\dagger}_{X,\downarrow,-k}]^{T}$ with $X=A,B$.
\begin{align}
    H&^\text{PMSC}(k)=\nonumber\\
    &\small \begin{bmatrix}
        h_{AA}(k) & \Delta(k) & h_{AB}(k) & 0\\
        -\Delta^{*}(-k) & -h_{AA}^{*}(-k) & 0 & -h_{AB}^{*}(-k)\\
        h_{BA}(k) & 0 &  h_{BB}(k) &\Delta(k)\\
        0 & -h_{BA}^{*}(-k) & -\Delta^{*}(-k) & -h_{BB}^{*}(-k)
    \end{bmatrix},   
    \label{eq2}
\end{align}
with
\begin{align}
    h_{AA(BB)}(k)=&(-2t^{PM}\cos k_y-\mu)s_0 -(+) 2t_j^{PM}\cos k_y s_y,\\
    h_{AB(BA)}(k)=&-2t^{PM}\cos \frac{k_x}{2} s_0 -(+) 2 i t_j^{PM}\sin \frac{k_x}{2}s_x,
\end{align}
 where $\Delta(k)=\Delta_{0}is_ye^{\pm i\phi/2}$ represents the proximity-induced isotropic superconducting pairing in the PM, with $\Delta_0$ denoting the pairing amplitude and $\phi$ the macroscopic superconducting phase. The upper (lower) sign corresponds to the right (left) SC lead. The operator $\psi_{A/B,\uparrow/\downarrow}^{\dagger}$ creates an electron at `A' or `B' sublattice with spin `up' or spin `down' in PMSC. The parameters $t^{PM}$ and $\mu$ denote the hopping amplitude and chemical potential, respectively. The term $t_j^{PM}$ represents the spin-dependent hopping of electrons mediated by magnetic atoms, thus capturing the exchange field generated by them. This spin dependent modulation leads to the $p$-wave symmetric spin splitting of the electronic bands. Moreover, $s_i$ with $i=0,x,y,z$ are Pauli matrices acting on the spin space. Next, we write the BdG Hamiltonian for AM as
 
\begin{align}
    \hat{H}^{AM}=&\sum_k \Phi^{\dagger}_k H^{AM}(k) \Phi_{k}^{T},
    \label{eqq4}
\end{align}
here $\Phi_{k}=[\phi_{\uparrow,k},\phi_{\downarrow,k},\phi_{\uparrow,-k}^{\dagger},\phi_{\downarrow,-k}^{\dagger}]^{T}$ and the operator $\phi_{\uparrow/\downarrow,k}^{\dagger}(\phi_{\uparrow/\downarrow,k})$ creates (annihilates) an electron with spin `up' or spin `down' in AM, the form of $H^{AM}(k)$ is
\begin{align}
    H^{AM}(k)=\begin{bmatrix}
        h(k) & 0\\
        0 & -h^{*}(-k)
    \end{bmatrix},
    \label{eq5}
\end{align} 
with h(k) written as,
\begin{align}
\nonumber
    h(k)=& -2t^{AM}(\cos k_x+\cos k_y)s_0 -\mu s_0 -Us_0\\
    \nonumber
    &-2t^{AM}_j(\cos k_x-\cos k_y)\cos(2\alpha)s_z\\ \nonumber & +2t^{AM}_j  (\sin k_x \sin k_y )\sin(2\alpha)s_z\\
    \nonumber
    &+\lambda[(\sin k_{y} \cos{\alpha}-\sin k_{x}\sin{\alpha})s_{x}
    \\ 
&-(\sin k_{x}\cos{\alpha}+\sin k_{y}\sin{\alpha}) s_{y}], \label{eq4}
\end{align}
here $t^{AM}$, $t^{AM}_j$, and $\lambda$ are the hopping amplitude, the exchange field strength, and the Rashba SOC, respectively, in the AM region. We assume the same chemical potential, $\mu$, throughout the junction and gate potential $U$ applied to the AM region only.\\      
To compute the current, we use the expectation value of time derivative of number operator ($\hat{N_L}$) in left SC. Since the current is conserved, it would be the same throughout the JJ [\onlinecite{PhysRevB.109.024517,PhysRevB.110.014518}] and written as

\begin{align}
    \big<I\big>&=e ~\Big< \frac{d\hat{N_L}}{dt}\Big>=\frac{ie}{\hbar}\Big<[\hat{H}^{JJ},\hat{N_L}] \Big>,\nonumber\\
    &=-\frac{e}{h}\int dE \text{ Tr}[\Gamma_{z}{H^{C_L}}G^{<}_{LJ}(E)+\text{H.c.}].
    \label{5}
\end{align}

The lesser Green's function is computed using the relation $G^{<}_{LJ}(E)=-f(E)[G^{r}_{LJ}(E)-G^{a}_{LJ}(E)]$ where, $f(E)$ is the Fermi-Dirac distribution function for energy E and $G^{a(r)}_{LJ}$ is the advance (retarded) non-local Green's function  which in turn can be found using the recursive algorithm, which is explained in Appendix [\ref{appB}].\\

\begin{figure}[b]
\begin{tabular}{c c}
    \centering
    \includegraphics[width=0.5\linewidth]{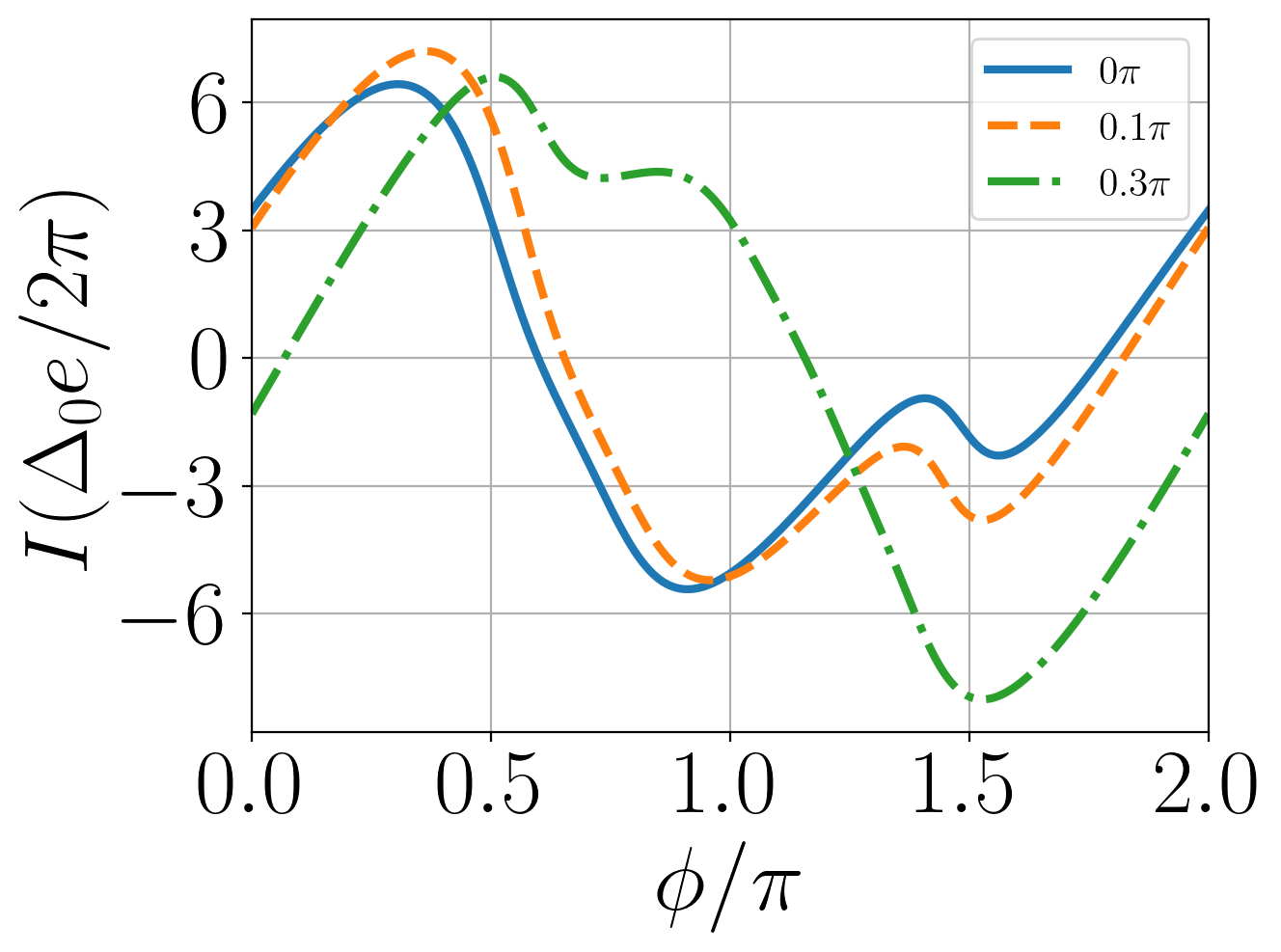}
    &\includegraphics[width=0.5\linewidth]{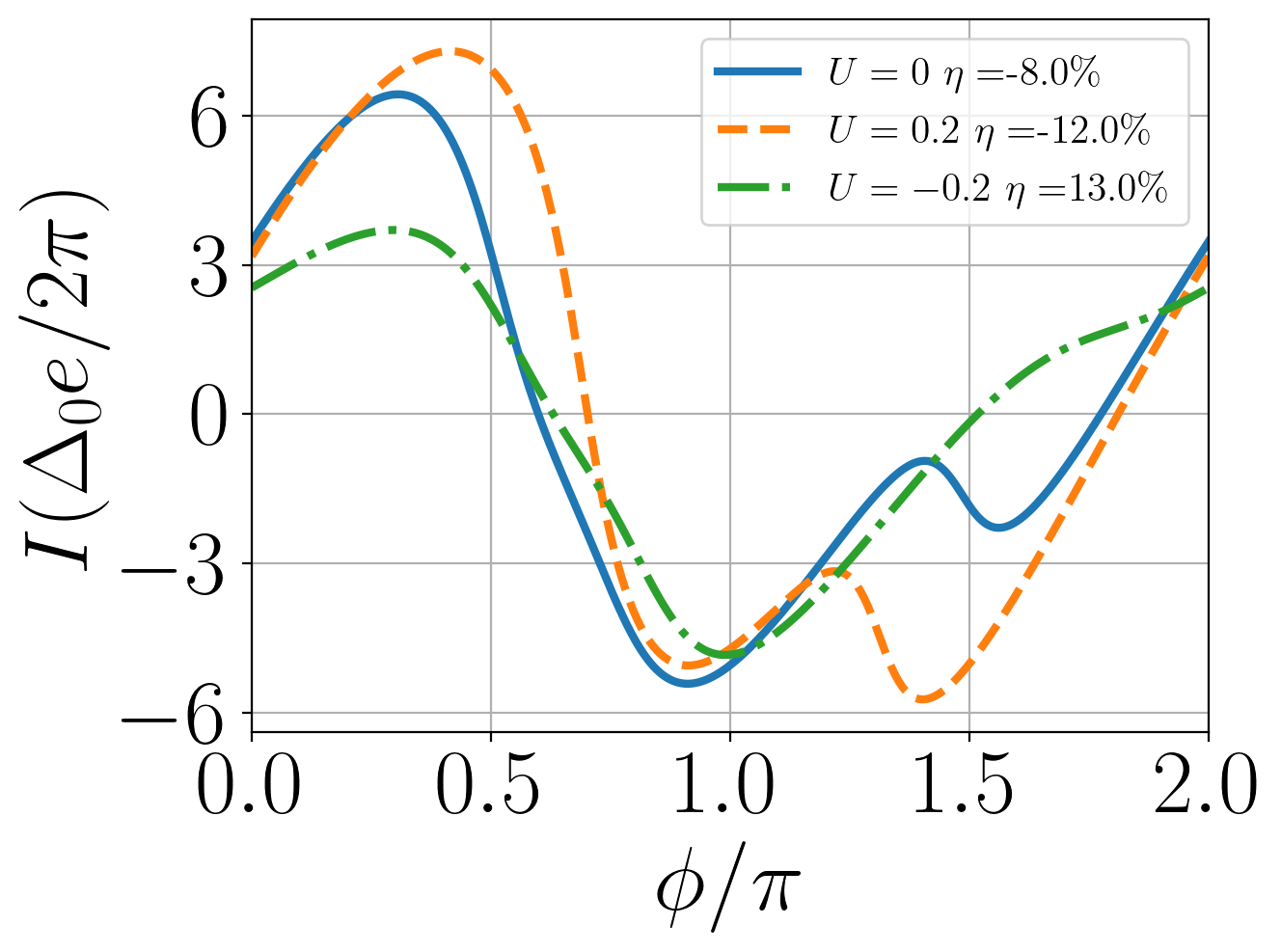}
    \end{tabular}
    \caption{(a) CPR plots for different crystallographic angles of AM, $\alpha=0$ (blue), $0.1\pi$ (orange), and $0.3\pi$ (green), demonstrating nonreciprocal critical currents ($|I_C^{+}|\neq|I_C^{-}|$) . (b) CPR plots for different gate potentials in the AM region at $\alpha=0$: $U=0$ (blue), $U=0.2$(orange) and $U=-0.2$(green). Other system parameters are $t_j^{PM}=0.35$, $t_j^{AM}=0.2$,  $t_0=1$, $\mu=-2$, and the length of the left and right SCs are taken as $N_x^{L(R)}=80$. The junction width is fixed at $N_{y}=6$ and the AM barrier length is $L_{x}^{AM}=5a$}.
    \label{fig2}
\end{figure}


\section{Results}
\label{sec3}
In this section, we present the numerical result for the current using Eq. (\ref{5}). As the two SC leads are coupled through an AM barrier, the current explicitly depends on the macroscopic phase difference $\phi$ and the crystallographic orientation of the AM.  The system parameters have also been taken as $\Delta=0.05t_{0}$, where $t_0=t^{AM}=t^{PM}$ and the barrier length as $L_{x}^{AM}=5a$, hence the system is in the short junction regime. For simplicity, we set the lattice constant to $a=1$ throughout the junction. Furthermore, we take $t_c=t_0$ and assume fully transparent left and right interfaces. In Fig.[\ref{fig2}], we plot the current phase diagram for different values of the crystallographic lobe angle of AM (Fig.[\ref{fig2}(a)]) and gate potential in AM region (Fig.[\ref{fig2}(b)]).
 Fig. [\ref{fig2}(a)] illustrates that the positive and negative values of the critical currents are different; hence, there is a finite diode effect in the junction. Notably, for $\alpha=0$ and $0.1\pi$, the current–phase relation (CPR) indicate a significant contribution from higher harmonics beyond the fundamental sinusoidal component. Fig. [\ref{fig2}(b)] demonstrate that CPR is tunable with the gate potential, such that it can also change the polarity of diode. For the rest of the paper, we set $U=0$ without loss of generality. These results show that despite using the same SC leads on both sides, we get non-reciprocity in the current, whereas different SCs or different crystallographic angles are among the necessary requirements in the previous models showing diode effect using AM [\onlinecite{yqsg-xdg8}]. 
 Moreover, Rashba spin–orbit coupling (SOC) is not required to realize the diode effect in this system. This is because the exchange-field–induced hopping itself breaks inversion symmetry, as will be demonstrated in the next section.


\begin{figure}[t]
\centering
\begin{tabular}{c c}
   \includegraphics[width=0.49\linewidth]{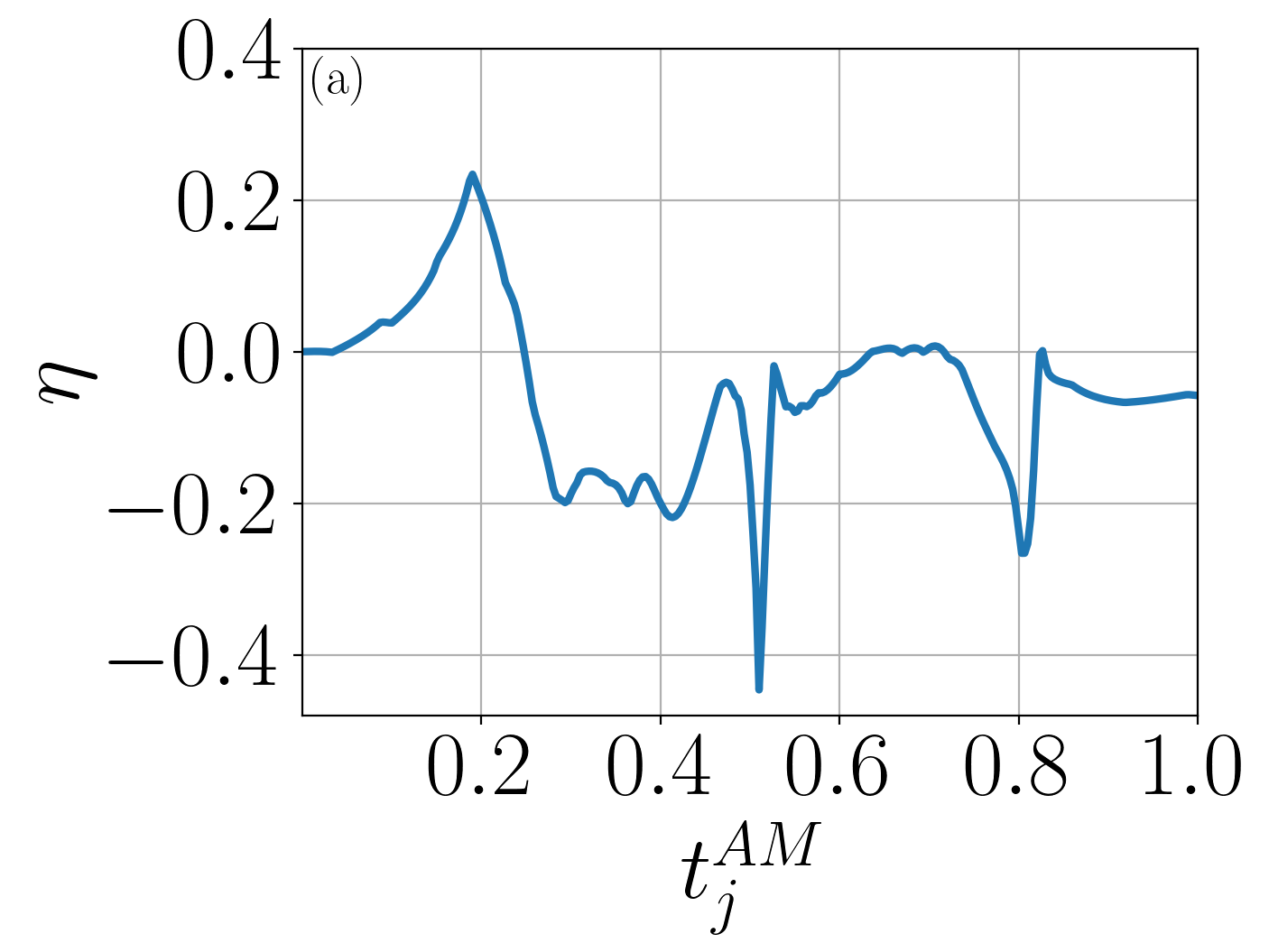}
& \includegraphics[width=0.5\linewidth]{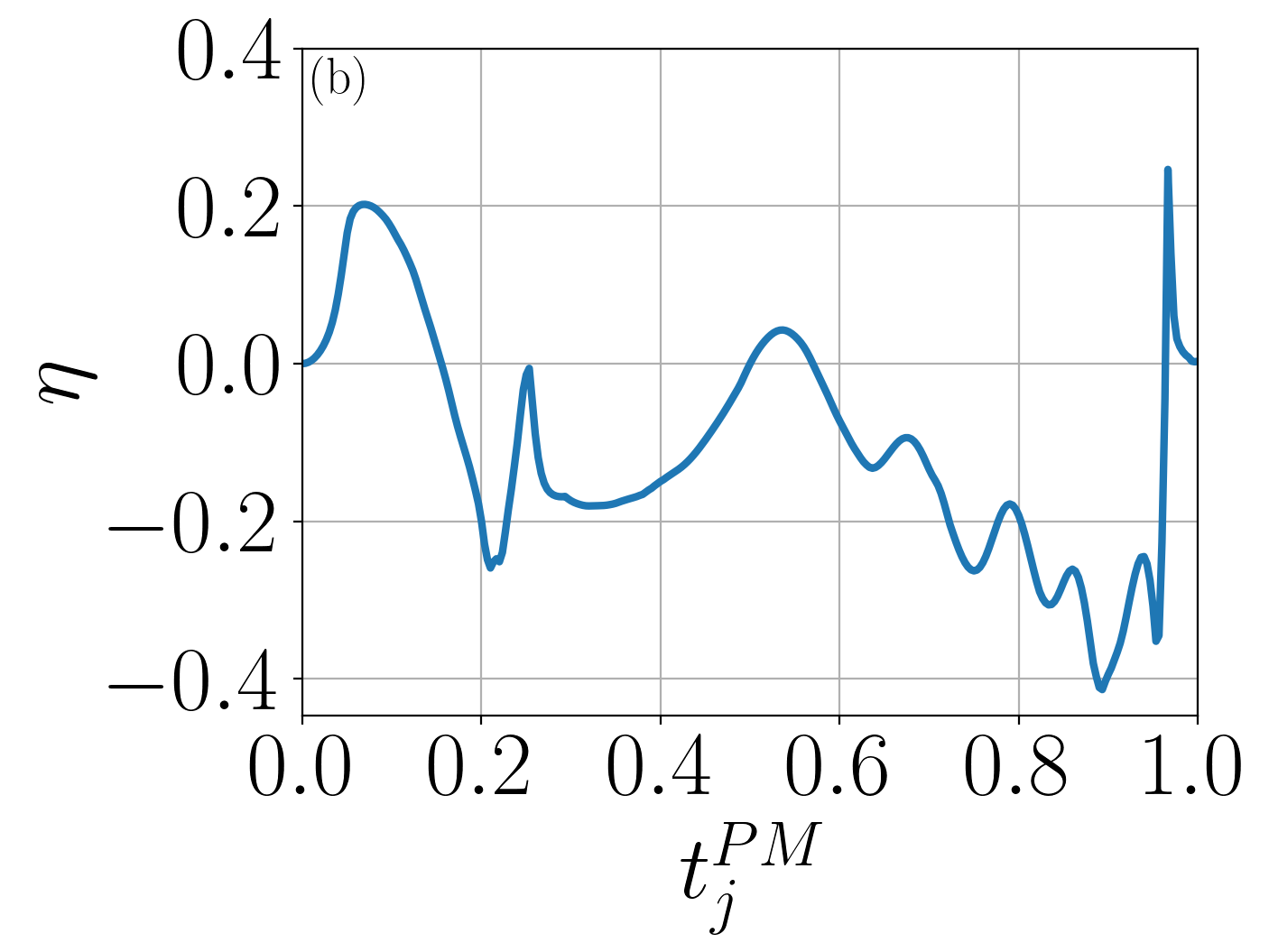}\\
\includegraphics[width=0.49\linewidth]{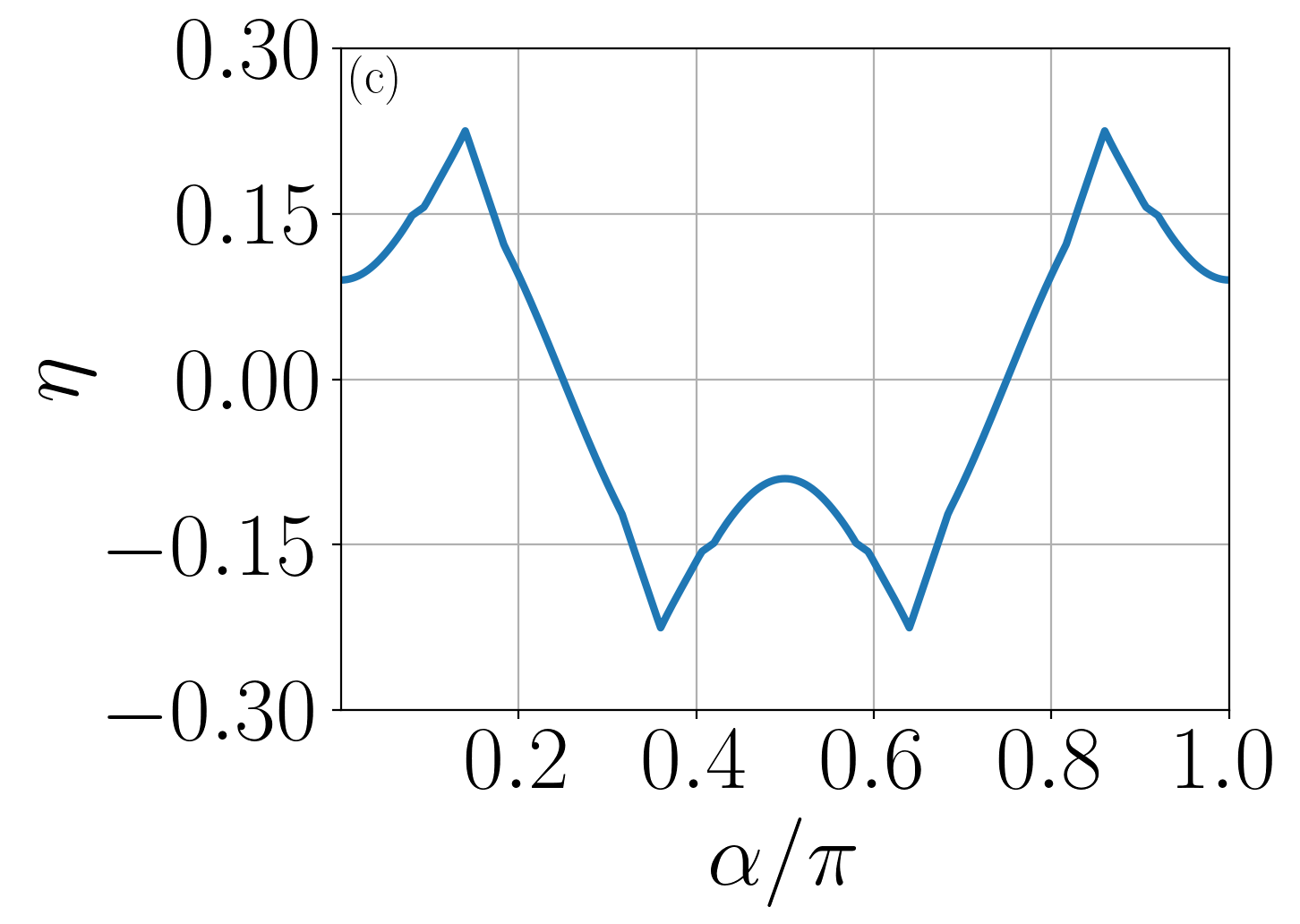}& \includegraphics[width=0.5\linewidth]{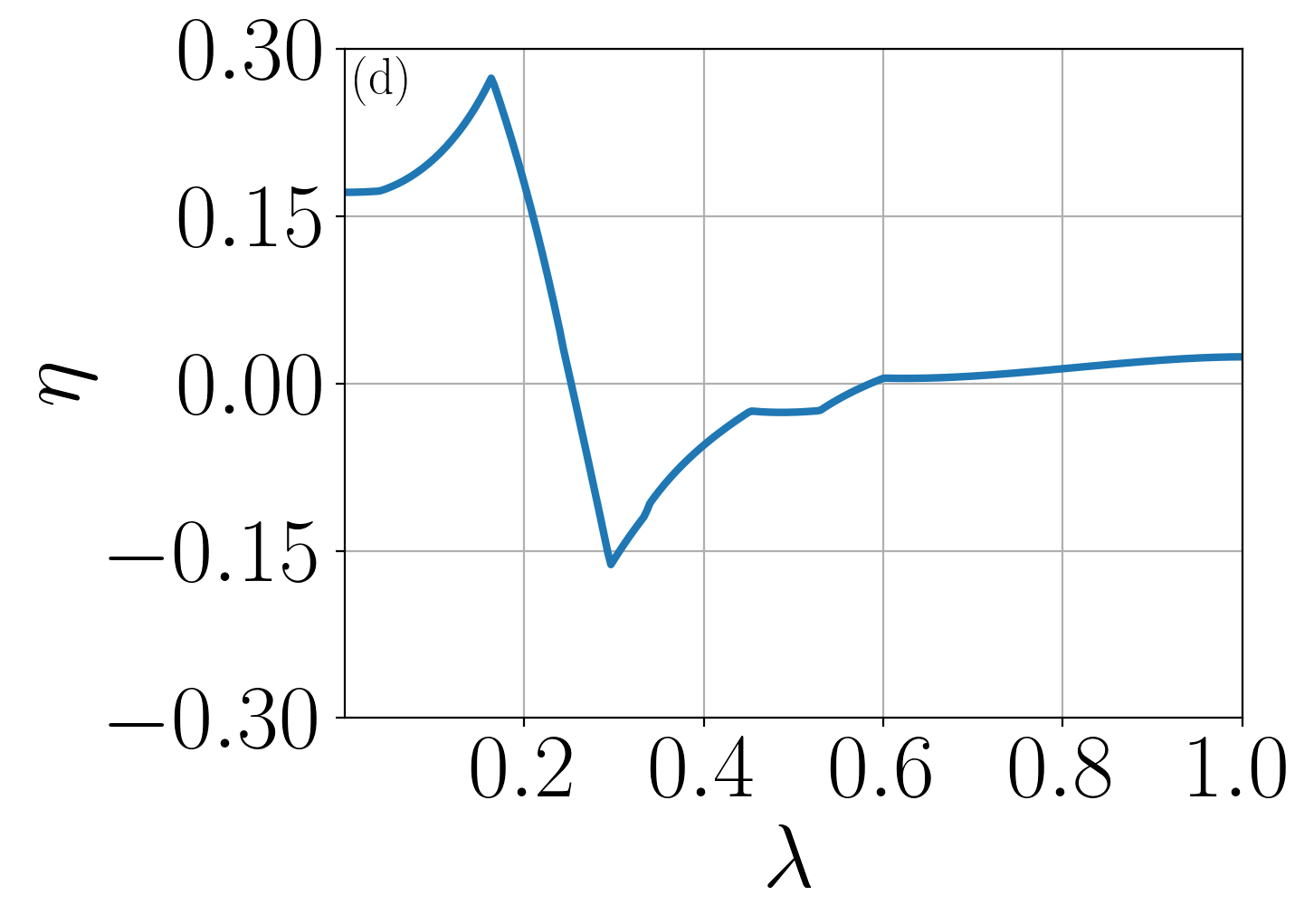}
    \end{tabular}
    \caption{(a-d) Variation of $\eta$ with (a) AM exchange field strength, $t_j^{AM}$, (b) PM exchange field strength, $t_{j}^{PM}$, (c) crystallographic lobe angle of AM, $\alpha$ and (d) Rashba Strength, $\lambda$. We consider $N_{x}^{L/S}=80$, $N_{x}^{AM}=5$ and $N_{y}=6$. In (a), (b), and (d), $\alpha=0$ and $\lambda=0$ for (a-c). $t_{j}^{PM}=0.2$ is set for (a), $t_{j}^{AM}=0.4$ for (b), $t_{j}^{PM}=0.35$ and $t_{j}^{AM}=0.2$ for (c) and $t_{j}^{PM}=0.1$ and $t_{j}^{AM}=0.4$ for (d)}
\label{fig3} 
\end{figure}
\begin{figure*}
\centering
\begin{tabular}{c c c}
   \includegraphics[width=1.0\linewidth]{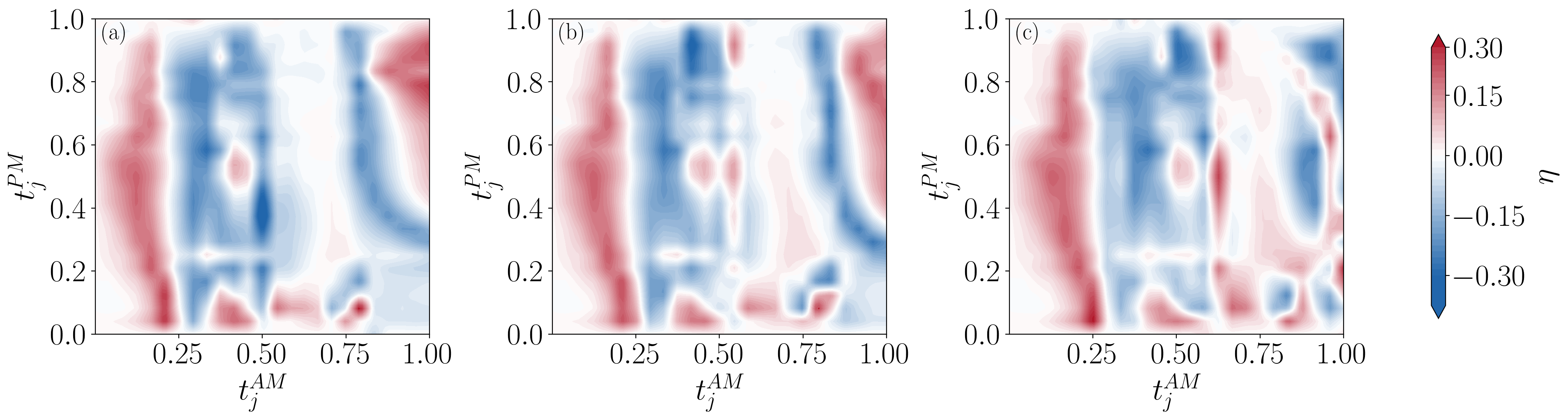}\\
   \includegraphics[width=1.0\linewidth]{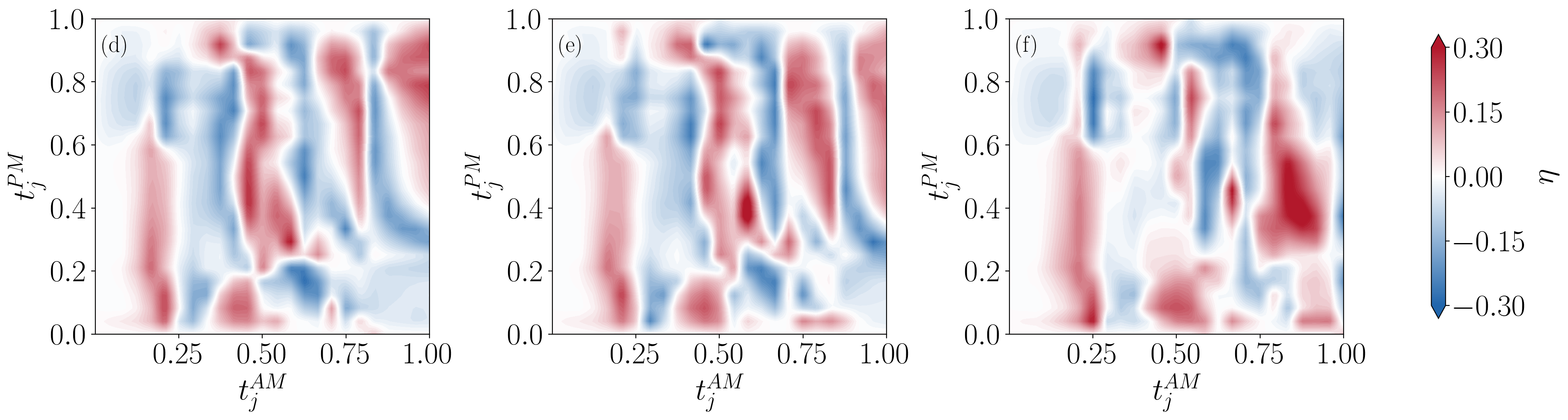}
    \end{tabular}
    \caption{(a-c) are contour plots of $\eta$(as colorbar) with $t_j^{AM}$ and $t_{j}^{PM}$ for both SC of ABAB.. configuration. (d-f) are corresponding contour plots for ABAB.. and BABA.. configuration of SC leads. Other parameters are $\alpha=0$ for (a,d), $\alpha=0.05\pi$ for (b,e), $\alpha=0.1\pi$ for (c,f), $N_{x}^{L/R}=80$, $N_{x}^{AM}=5$, $N_{y}=6$ and $\lambda=0$. }
\label{fig4} 
\end{figure*}

We quantify nonreciprocity using the efficiency parameter, given by $\eta=(|I_C^{+}|-|I_C^{-}|)/(|I_C^{+}|+|I_C^{-}|)$. Here, $I_C^{+}$  and $I_C^{-}$ denote the maximum value of $I(\phi)$ and $-I(\phi)$, respectively. Using this definition, we compute the efficiency $\eta$ as a function of $t_{j}^{PM}$, $t_{j}^{AM}$ and $\alpha$ as shown in Fig. \ref{fig3}(a-c) with $\lambda=0$. However, the effect of Rashba SOC in the AM region is shown in Fig. \ref{fig3}(d). The dependence of $\eta$ on $t_{j}^{AM}$ (Fig.\ref{fig3}(a)) shows that high efficiency is maintained over a wide interval for $t_{j}^{AM}<0.5$, providing a broad and stable window of high diode efficiency.  This behavior remains qualitatively unchanged upon varying $\alpha$ and $t_{j}^{PM}$. In contrast, the variation of efficiency with $t_j^{PM}$ (Fig. \ref{fig3}(b)) shows that the fluctuation becomes more pronounced as $t_j^{PM}$ increases, Notably, there are broad ranges of $t_j^{PM}$ for which efficiency is significantly high. The dependence of $\eta$ on $\alpha$ (Fig. [\ref{fig3}(c)]) shows an interesting angular response. Over the broad range, $\eta$ varies approximately linearly with $\alpha$, even though the relevant terms enter the Hamiltonian through sinusoidal functions (Eq. \ref{eqq4}). Notably, $\eta$ vanishes at $\alpha=\pi/4$, this feature follows from symmetry considerations. Now, to analyze the role of SOC in AM region, we plot efficiency variation with $\lambda$ (Fig.[\ref{fig3}(d)]). Although inversion is already broken in our setup, introducing Rashba SOC introduces an additional inversion-asymmetric SOC contribution that splits the spin degeneracy in AM and enhances the splitting. Consequently $\eta$ is enhanced for moderate value of $\lambda$, indicating stronger non-reciprocity in CPR. However, for large value of $\lambda$, $\eta$ becomes less fluctuating and can drop below the value obtained in the absence of Rashba SOC. Moreover Rashba SOC can be used to reverse the polarity of efficiency in CPR. 

We next present the contour plots of $\eta$ as a function of $t_j^{AM}$ and $t_j^{PM}$ for $\alpha=0$, $0.05\pi$, and $0.1\pi$, shown in Figs. [\ref{fig4}(a–c)], respectively. These maps reveal a clear trend: for a fixed $t_j^{AM}$, the efficiency remains high over a broad range of $t_j^{PM}$, consistent with the behavior observed in the one-dimensional dependence shown in Fig. [\ref{fig3}(b)]. Importantly, for all values of $\alpha$, a high efficiency $\eta$ persists over a broad range of $t_j^{AM}$ and $t_j^{PM}$. This robustness indicates that the nonreciprocal response is not fine-tuned to a specific lobe angle, making it well suited for practical implementations without stringent constraints on the exchange field strengths. We also consider an alternative configuration for the SC leads. The geometry of the left SC lead shown in Fig. [\ref{fig1}(a)] corresponds to an ABAB.. stacking, where `A’ and `B’ denote consecutive 1D strips of the two sublattices. An equally valid choice is the BABA.. stacking, in which the lattice begins with the `B’ sublattice on the left and terminates with the `A’ sublattice. Accordingly, we evaluate the efficiency for a junction composed of ABAB and BABA stackings in the left and right SC leads, respectively. Figs. [\ref{fig4}(d–f)] present the corresponding contour maps of  $\eta$ in the $t_j^{PM}$–$t_j^{AM}$ plane for the ABAB/AM/BABA configuration. Overall, the qualitative behavior of $\eta$ closely resembles that of the ABAB/AM/ABAB junction shown in Figs. [\ref{fig4}(a–c)], although with more pronounced fluctuations. In both configurations, $\eta$ reaches values of up to around 45\% , and regions with $\eta > 20\% $ span a substantial portion of the parameter space, demonstrating that the high-efficiency regime is robust against variations in the SC lead configuration.

\section{Symmetry Analysis}
\label{sec4}
We study several symmetry transformations of the JJ Hamiltonian to analyze the role of individual terms and their effect on the behavior of $\eta$ discussed in Sec. \ref{sec3}. In particular, we define time reversal ($\mathcal{T}$), mirror reflection in $xz$ plane ($M_{xz}$), mirror reflection in $yz$ plane ($M_{yz}$), $\pi$ rotation of spin about $y-$axis ($R_y$) and $z-$axis ($R_z$). The annihilation operator transforms under these symmetry operations as follows,
\begin{align}
    \nonumber
    \mathcal{T}c_{{\bf k},\delta,s}\mathcal{T}^{-1}=& s c_{-{\bf k},\delta,\bar{s}}
    \\
    \nonumber
    M_{xz}c_{{\bf k},\delta,s}M_{xz}^{-1}=& \bar{s} c_{(k_x,-k_y),\delta,\bar{s}}
    \\
    \nonumber
    M_{yz}c_{{\bf k},\delta,s}M_{yz}^{-1}=& -i c_{(-k_x,k_y),\bar{\delta},\bar{s}}
    \\
    \nonumber
    R_{y}c_{{\bf k},\delta,s}R_y^{-1}=& s c_{{\bf k},\delta,\bar{s}}
    \\
    R_zc_{{\bf k},\delta,s}R_z^{-1}=& is c_{{\bf k},\delta,s}
\end{align}
Here, ${\bf k}=(k_x,k_y)$ is the momentum, while the subscripts $\delta (\bar{\delta})=A/B(B/A)$ and $s (\bar{s})=\uparrow/\downarrow(\downarrow/\uparrow)$ represent sublattice (only for PMSC) and spin degree of freedom, respectively. For the AM region, the sublattice degree of freedom can be ignored. The matrix representation of these symmetry transformations is given in Appendix [\ref{app C}]. 

Since the JJ is formed along the $x-$axis and we are interested in the longitudinal current, we first examine how the mirror operation $M_{yz}$ acts on the junction geometry. Let the macroscopic superconducting phases of the left and right PMSC leads be $\phi_{L}$ and $\phi_{R}$, respectively, and define the phase difference as, $\phi=\phi_{L}-\phi_{R}$. Under the mirror operation 
$M_{yz}$, the spatial coordinate along the junction is inverted, causing the two PMSC leads to exchange their positions: the left lead is mapped onto the right lead and vice versa. Importantly, the intrinsic Hamiltonian of the PMSC remains unchanged under this mirror operation; only the phases associated with the spatially separated leads are interchanged. Consequently, the PMSC part of the JJ Hamiltonian transforms as follows:
\begin{align}\nonumber
    M_{yz}H^{PMSC}(\phi_{L(R)},t^{PM},\mu,t_j^{PM})M_{yz}^{-1}=\\H^{PMSC}(\phi_{R(L)},t^{PM},\mu,t_j^{PM}).
    \label{eq19}
\end{align}
Thus, $M_{yz}$ maps the junction configuration with phase bias $\phi$ to an equivalent configuration with phase bias $-\phi$ upto the $M_{yz}$ transformed barrier region. In the AM region, $M_{yz}$ acts as
\begin{align}\nonumber
    M_{yz}H^{AM}(\alpha, t^{AM},\mu,t_j^{AM})M_{yz}^{-1}=\\H^{AM}(\pi/2-\alpha ,t^{AM},\mu,t_j^{AM}).
    \label{eq20}
\end{align}
This implies that the AM Hamiltonian is not invariant under $M_{yz}$ transformation at fixed $\alpha$, instead, $M_{yz}$ maps $\alpha$ to $\pi/2-\alpha$. Therefore, the transformation $M_{yz}$ in Eqs. (\ref{eq19},\ref{eq20}) implies the current $I(\alpha, \phi)=-I(\pi/2-\alpha,-\phi)$, since $M_{yz}$ also reverses the current direction along the $x$-axis. This relation can be used to explain Fig. [\ref{fig3}(c)], where $\eta(\alpha)=-\eta(\pi/2-\alpha)$ can be seen.

\begin{table}[ht]
\centering
\begin{tabular}{>{\centering\arraybackslash}p{3cm} >{\centering\arraybackslash}p{1.2cm} >{\centering\arraybackslash}p{1.2cm} >{\centering\arraybackslash}p{1.2cm} >{\centering\arraybackslash}p{1.2cm} >{\centering\arraybackslash}p{1.2cm}}
    \toprule
    & TRS & Mxz & Myz & Rz & Ry \\
    \midrule
    NM & \ding{51} & \ding{51} & \ding{51} &  \ding{51} & \ding{51} \\
    $-2t_{j}^{PM}\sin (k_{x}/2)\rho_{y}\tau_{z}s_{x}$ & \ding{56} & \ding{56} & \ding{51}  & \ding{56} &\ding{56} \\
    $-2t_{j}^{PM}\cos k_{y} \rho_{z}\tau_{0}s_{y}$ & \ding{56} & \ding{51} & \ding{51} &  \ding{56} &\ding{51} \\
    $d_{x^2-y^2} \tau_z s_{z}$ & \ding{56} & \ding{56} & \ding{56}  & \ding{51} &\ding{56}\\
    $d_{xy}\tau_z s_{z}$ & \ding{56} & \ding{51} & \ding{51}  & \ding{51}&\ding{56}\\
    $m_{z}\tau_z s_{z}$ & \ding{56} & \ding{56} & \ding{56}  & \ding{51} &\ding{56}  \\
    $\hat{\Delta}(\phi)$  & \ding{56}($\phi\rightarrow-\phi$) & \ding{56}($\phi\rightarrow-\phi$) & \ding{51}  & \ding{51}&\ding{51} \\
    \bottomrule
\end{tabular}
\caption{Transformation of each terms of the Hamiltonian of JJ under TRS, $M_{xz}$, $M_{yz}$ and the rotation $R_{z}$. Here, NM denotes the normal state hopping and chemical potential terms in both SC leads and barrier region. $d_{x^2-y^2}$ and $d_{xy}$ are altermagnet exchange field for $\alpha=0$ and $\alpha=\pi/4$ respectively and $\hat{\Delta}$ represents pairing potential of $s-$wave SC. The symbols \ding{51} and \ding{56} indicate, respectively, that a term is invariant or acquires an overall minus sign under the corresponding operation except for $\hat{\Delta}$, where it represent $\phi\rightarrow-\phi$}
\label{tab1}
\end{table}

However, at $\alpha=\pi/4$, mirror symmetry constrains the CPR such that $I(\phi)=-I(-\phi)$. Consequently, any positive current at a given phase bias is paired with a negative current of equal magnitude at the opposite phase bias, which leads to $\eta=0$. Thus, despite the broken IS (in PMSC) and TRS (in both the AM and PMSC)(see Table[\ref{tab1}]), the diode effect is absent. Whereas, for $\alpha \neq \pi/4$, the CPR become non-reciprocal as shown in previous section. Thus, beyond TRS and inversion, the mirror operation $M_{yz}$ remains the decisive symmetry constraint. In the present junction it is broken by the AM (Eq.[\ref{eq20}]), while inversion symmetry remains preserved in the AM.  Next, we highlight the role of the PM exchange field $t_{j}^{PM}$ in enabling the JDE. To this end, we introduce two combined symmetry operations, $\mathcal{X}=\mathcal{T}R_y$ and $\mathcal{Y}=M_{xz}M_{yz}$. As summarized in Table[\ref{tab1}], the two PM-induced terms show a complementary symmetry character: the term  $\propto \sin(k_x/2)$ violates $\mathcal{X}$ but preserves $\mathcal{Y}$, whereas the term $\propto \cos k_y$ violates $\mathcal{Y}$ but preserves $\mathcal{X}$. Therefore, both terms are necessary for the diode effect, since the AM respects both symmetries and enforces the constraint $I(\phi)=-I(-\phi)$ in their absence. This behavior is evident in Fig. [\ref{fig3}(a,b)] where $\eta\rightarrow0$ as $t_j^{PM}$ tends to zero.

So far, we have focused on the case where both the SC leads have same ABAB configuration. We now consider the case when left and right leads realize the ABAB and BABA configuration, respectively. The argument follows also applies to the vice versa case. In this case, we introduce the combined operation $\mathcal{Z}=M_{yz}R_{z}$ acting on the junction geometry, which transform AM region Hamiltonian in the same way as in Eq. [\ref{eq20}]. However, the intrinsic Hamiltonian of PMSC transforms as, 
\begin{align}\nonumber
    \mathcal{Z}H_{AB(BA)}^{PMSC}(\phi_{L(R)},t^{PM},\mu,t_j^{PM})\mathcal{Z}^{-1}=\\
    H_{AB(BA)}^{PMSC}(-\phi_{R(L)},t^{PM},\mu,t_j^{PM}).
    \label{eq12}
\end{align}
\begin{figure}[t]
    \includegraphics[width=0.9\linewidth]{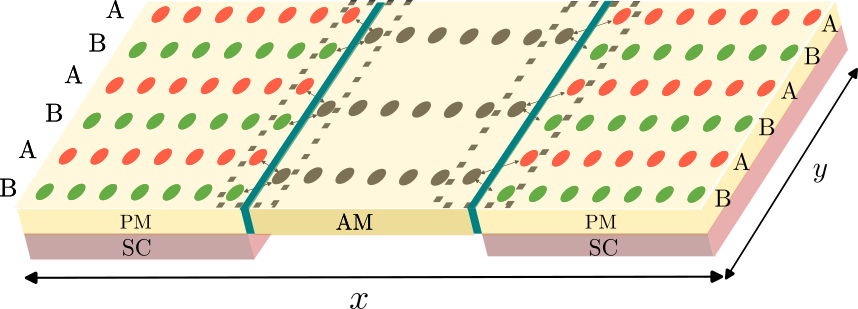}
    \caption{Schematic of a Josephson junction comprising $90^{0}$ rotated PMSC leads separated by an AM barrier. At the left interface, both A and B sublattices of the PMSC are coupled to a nearest single AM lattice site.[\onlinecite{Fukaya2025}]}
    \label{fig5a}
  \end{figure}
Here, we define Hamiltonian corresponding to the ABAB and BABA configuration as, $H^{PMSC}_{AB}$ and $H^{PMSC}_{BA}$ respectively. Hence, the symmetry constraint implied by Eq. [\ref{eq12}] is the same as that implied by Eq. [\ref{eq19}] and remains valid for any analogous transformation in which $M_{yz}$ is replaced by 
$\mathcal{Z}$ in this hybrid geometry.

\section{JDE with other Minimal Models, Orientations, and Ferromagnet as barrier}

\begin{figure}[b]
\begin{tabular}{c c}
    \centering
    \includegraphics[width=0.5\linewidth]{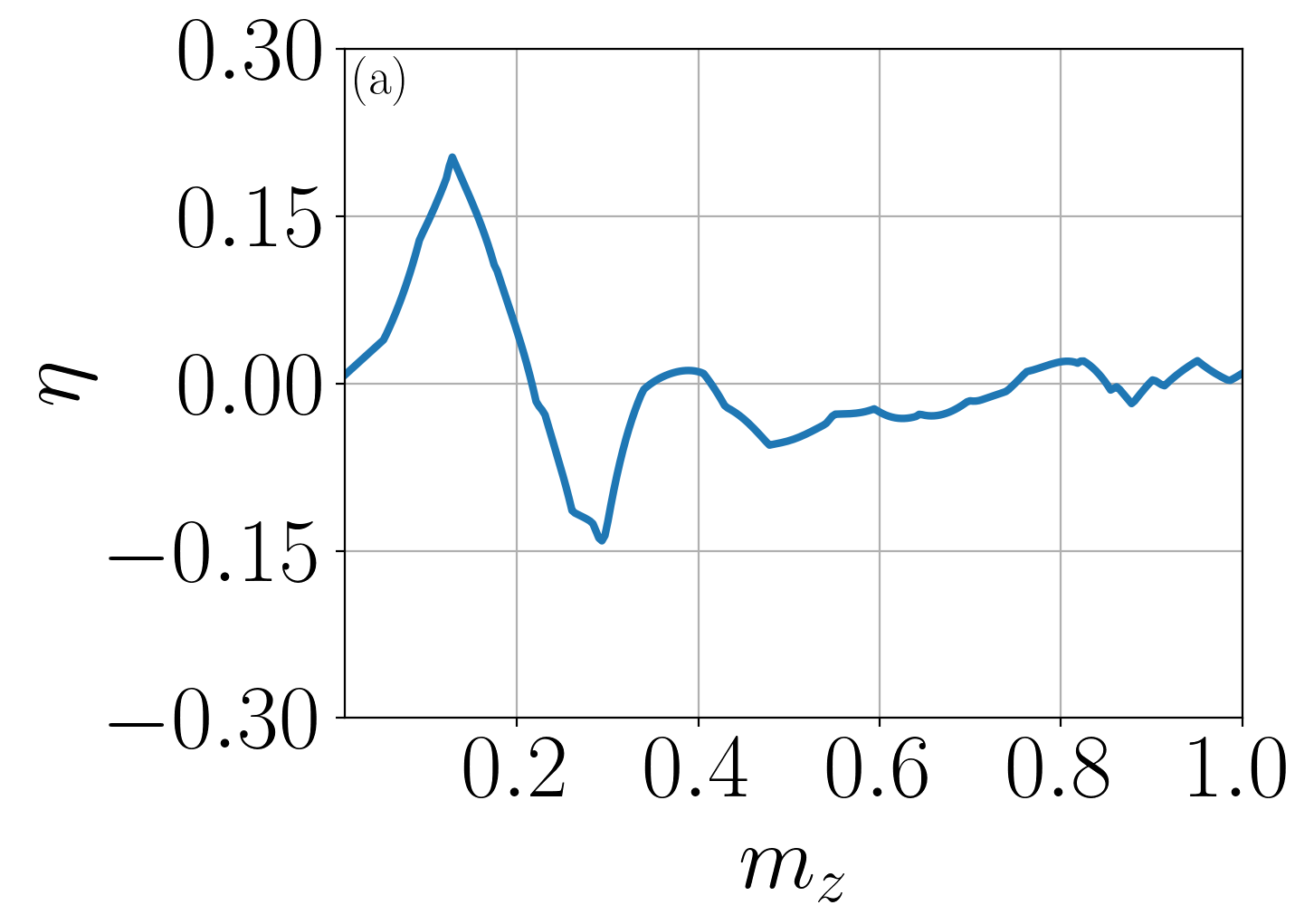}
    &\includegraphics[width=0.5\linewidth]{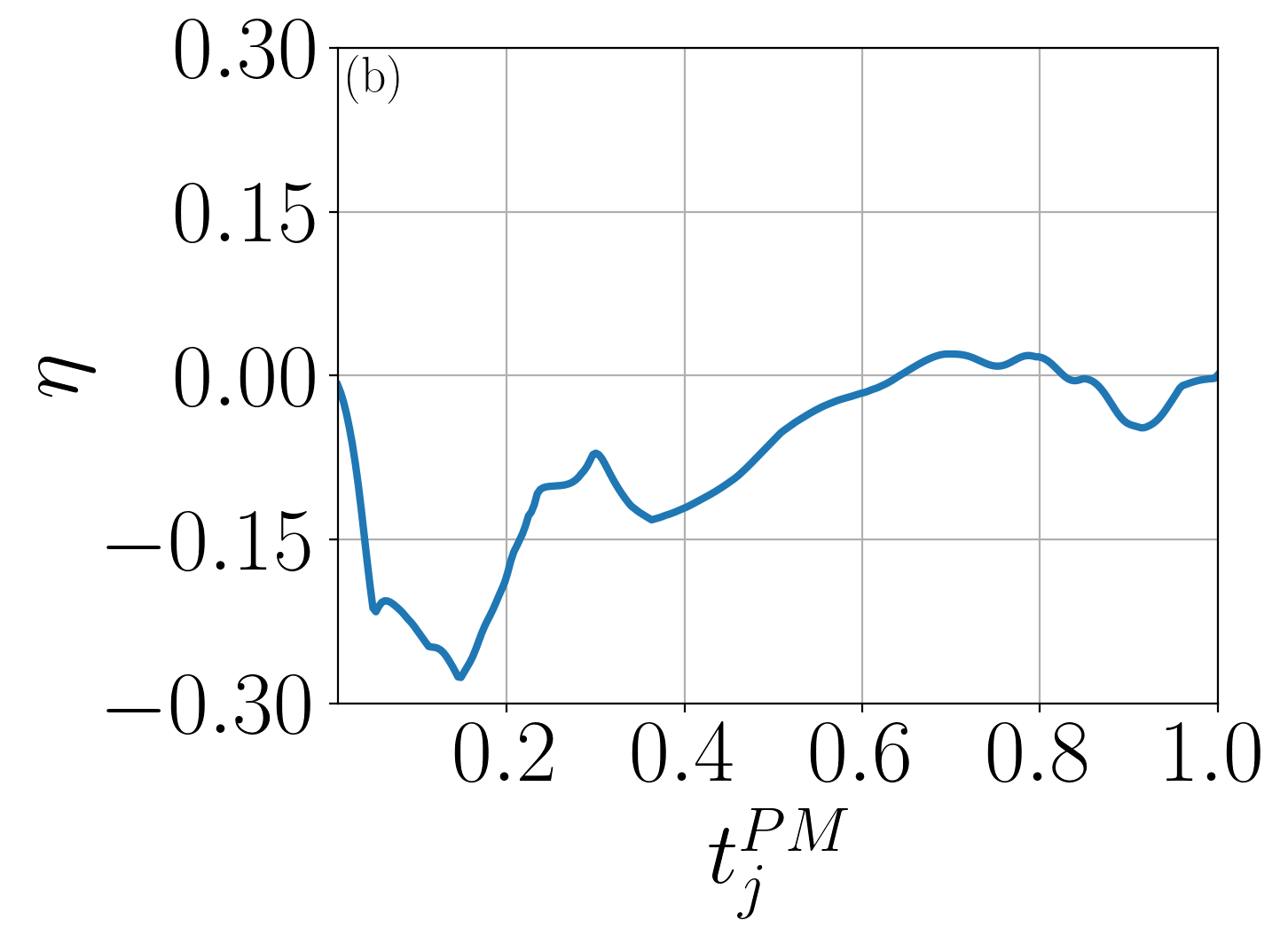}
    \end{tabular}
    \caption{(a) Variation of efficiency $\eta$  with ferromagnetic strength $m_z$, where $t_j^{PM}=0.35$ and (b) Variation of efficiency $\eta$ with PM exchange field ${t_j}^{PM}$, where $m_z=0.3$. Rest of the parameters are same as in Fig.[\ref{fig3}].}.
    \label{figg 5}
\end{figure}

\subsection{Other minimal models and orientation of junction}
Since the $p$-wave magnetism is protected by the combined operation of time reversal and lattice translation, the effective electronic Hamiltonian in such systems can be formed in different ways depending on the underlying magnetic lattice structure. Accordingly, here we also consider another minimal model for $p$-wave magnet [\onlinecite{PhysRevLett.133.236703}] with such properties in this section. The effective Hamiltonian for the PM is given as,
\begin{align}
\nonumber
    H=&-\{2t(\cos{ k_xa} + \cos{k_ya}) +\mu \}\sigma_{0}\tau_{0} \\
    \nonumber& +(\alpha_{x}\sin{k_xa} + \alpha_{y}\sin{k_ya})\sigma_{z}\tau_{0}\\
    & + J_{sd}\sigma_{x}\tau_{z} + 2t_{int}\cos{(k_xa/2)\sigma_{0}\tau_{x}}
\end{align}

Here $\sigma_{i}$ and $\tau_{i}$ are Pauli matrices acting on spin and sublattice space. ``t" is intrasectoral(sublattice) hopping, ``$J_{sd}$" is sd coupling within sector and ``$t_{int}$" is intersectoral hopping. We construct the JJ using this model in same geometry as in Fig.[\ref{fig1}(a)], where sublattices are stacked along the junction. Notably, in this case also, we find the diode effect even without including Rashba SOC in the barrier region. Moreover, we also compute the CPR for a junction in which the p-magnet is rotated by $90^0$, as shown in Fig.[\ref{fig5a}]. In this junction we used coupling among PMSC and AM in such a way that both $AB$ sublattice in PM are coupled to a sublattice point in AM, following the construction of Ref.[\onlinecite{Fukaya2025}] (see Eq.[\ref{A17}]). In this junction also CPR is found to be non-reciprocal in both the given models.
\subsection{Ferromagnet as a barrier in Junction}

 We next replace the AM barrier with a ferromagnetic (Fe) barrier in the same Josephson-junction geometry ([Fig\ref{fig1}(a)]). In Nambu basis the Hamiltonian for the ferromagnet region ${H^{Fe}}(k)$ is,
\begin{align}
    H^{Fe}(k)=\begin{bmatrix}
        h(k) & 0\\
        0 & -h^{*}(-k)
    \end{bmatrix},
\end{align} 
where, $h(k)= -2t^{Fe} (\cos k_x+\cos k_y  )s_0 -\mu s_0$+$m_z{s_3}$.
Similar to the AM case, Fe also exhibits spin split fermi surface  due to broken  TRS, however it is isotropic splitting.

We find that this JJ also exhibits non-reciprocity as shown in Fig. [\ref{figg 5}]. Fig. [\ref{figg 5}(a)] presents the variation of $\eta$ with the ferromagnetic strength, indicating a strong dependence of the efficiency on it. Similar to the AM case, $\eta$ remains high over a broad range for moderate exchange strengths ($m_z < 0.5$), but decreases substantially at larger $m_z$. In both the AM and Fe barrier junctions, the key additional ingredient that enables nonreciprocity is the breaking of mirror symmetry $M_{yz}$. However, in the absence of the AM or Fe barrier, $M_{yz}$ is preserved even though the IS and TRS are already broken in the overall junction and CPR is reciprocal. Thus, breaking $M_{yz}$ distinguishes the AM/Fe cases and allows $\eta\neq0$. 
To complete the analysis, we plot the variation of $\eta$ with PM exchange field Fig. [\ref{figg 5}(b)]. We find that $\eta$ decreases once the exchange field exceeds a moderate value ($t_{j}^{PM}=0.5$ in this case), indicating that stronger exchange does not necessarily enhance nonreciprocity in this configuration.

\section{Conclusion}
\label{sec5}
In this work, we examine the CPR of a JJ that incorporates UMs. Specifically, we consider a setup where proximity-induced superconductivity in a PM forms the superconducting leads, while an AM acts as the barrier. We find that the junction exhibits a pronounced diode effect, with an efficiency of up to $45\%$, even in the absence of an external magnetic field, thus achieving a field-free JDE. Remarkably, Rashba spin–orbit coupling is not required in this planar junction to break the IS and induce nonreciprocity, since IS is already broken in proximity-induced superconducting PM leads. We also note that the JDE can be achieved without employing different superconducting leads across the junction. In addition to TRS and IS, mirror symmetry operation $M_{yz}$ is the key ingredient that enables nonreciprocity. Through symmetry arguments, it has been shown that both the UMs are necessary in this model for field free JDE. 
Moreover, high efficiency is achieved over a wide range of system parameters, while the symmetry-imposed constraints are limited to specific orientations of the AM crystallographic axis. As a result, the number of conditions required to obtain high efficiency in this system is considerably smaller than in previous works employing AM-based planar Josephson junctions to realize a field-free JDE.
We also study two configurations of the PMSC leads, as well as other minimal models with a FE barrier, and compute the corresponding efficiencies. These results establish that UM-based JJs can sustain diode effects over wide parameter ranges in every configuration considered.

\appendix

\begin{widetext}
\section{Real Space JJ Hamiltonian Matrix}
\label{appA}
The discrete space (lattice) representation of the Hamiltonian of the junction can be found using the Fourier transformation of the $k$-space Hamiltonian of respective regions given in Eq.[\ref{eq2},\ref{eq5}]. We now rewrite PMSC-AM-PMSC junction in real space representation as,
\begin{align}
    \hat{H}^{JJ}=&\sum_{r,r'} \Lambda^{\dagger}_{r} H^{JJ}_{r,r'}\Lambda_{r'}
    \label{eq8}
\end{align}
Where, $\Lambda_{r}=[c_{r,\uparrow},c_{r,\downarrow},c_{r,\uparrow}^{\dagger},c_{r,\downarrow}^{\dagger}]^{T}$ and $r,r'\in (x,y),(x',y')$ are lattice points in the junction. The operator $c_{r,\uparrow/\downarrow}$ annihilates an electron at  position $r$ with spin `up' or `down'. 
Here, $\hat{H}^{JJ}$ can be expanded as,
\begin{align}
\hat H^{JJ}
&= \sum_{r,r'\in \mathcal{S}}
\Lambda_r^{\dagger}\,H^{PMSC}_{rr'}\,\Lambda_{r'}+
\sum_{r,r'\in \mathcal{A}}
\Lambda_r^{\dagger}\,H^{AM}_{rr'}\,\Lambda_{r'} + \hat{H}^{C}.
\label{eq9}
\end{align}
where $\mathcal{S}\equiv \mathcal{S}_L\cup\mathcal{S}_R$, $\mathcal{S}_{L(R)}=\{(x,y): x\le -a (x\ge (N_x^{AM}+1)a)\}$ and $\mathcal{A}=\{(i_x,i_y): 1\le i_x\le N_xa\}$ and $\hat{H}^{C}$ represent the coupling at left and right junction given as,
\begin{align}
\nonumber
    \hat{H}^{C}=&\Lambda^{\dagger}_{x=-a,y} H^{C_{L}}\Lambda_{x=0,y} + \\&\Lambda^{\dagger}_{x=N_x^{\small AM}+1,y} H^{C_{R}}\Lambda_{x=N_{x}^{\small AM}a,y} + H.c
    \label{eq10}
\end{align}

For PMSC region (both left and right), $H^{PMSC}_{r,r'}$ in Eq. [\ref{eq9}] can be written as,
\begin{align}
    H_{AA}^{PMSC}=H_{BB}^{PMSC}&=\begin{bmatrix}
        -\mu s_0 & \Delta(is_y) e^{\pm i\frac{\phi}{2}}\\
        -\Delta(is_y)e^{\mp i\frac{\phi}{2}} & \mu s_0
    \end{bmatrix}
    \label{eqA1}
\end{align}\\
where $  H_{AA}^{PMSC}$and $H_{BB}^{PMSC}$ are onsite energy on `A' and `B' sublattice. For hopping in $x-$direction,
\begin{align}
    H_{AB(BA)}^{PMSC}&=\begin{bmatrix}
        -t^{PM} s_0 -(+)t_j^{PM}s_x & 0\\
       0 & t^{PM} s_0 +(-)t_j^{PM}s_x
    \end{bmatrix}
\end{align}
where  $H_{AB(BA)}^{PMSC}$ is hopping matrix from `A'(`B') sublattice to `B'(`A') in $x-$direction.  For $y-$direction,
\begin{align}
    H_{AA_y(BB_y)}^{PMSC}&=\begin{bmatrix}
        -t^{PM} s_0 -(+)t_j^{PM}s_y & 0\\
       0 & t^{PM} s_0 -(+)t_j^{PM}s_y
    \end{bmatrix}
\end{align}
Since sublattice are same in $y-$direction, $H_{AA_y(BB_y)}^{PMSC}$ represent hopping in $y-$direction from `A'(`B') sublattice to `A'(`B') sublattice. Corresponding to these, strip Hamiltonian are defined as,

\begin{align}
    H_{AA(BB)}=
    \begin{bmatrix}
        H_{AA(BB)}^{PMSC} & {H_{AA_y(BB_y)}^{PMSC}} & & & & \\
        {H_{AA_y(BB_y)}^{PMSC}}^{\dagger} & H_{AA(BB)}^{PMSC}& {H_{AA_y(BB_y)}^{PMSC}} & & &\\
        & {H_{AA_y(BB_y)}^{PMSC}}^{\dagger} & H_{AA(BB)}^{PMSC}& {H_{AA_y(BB_y)}^{PMSC}} & & \\
        & & ...& &  &\\
        & & & ...& & {H_{AA_y(BB_y)}^{PMSC}}\\
        & & & &{H_{AA_y(BB_y)}^{PMSC}}^{\dagger} &  H_{AA(BB)}^{PMSC}
    \end{bmatrix}
\end{align}

For $H_{AA/BB}^{LS(RS)}$ use Eq. [\ref{eqA1}] with upper(lower) sign. For hopping to the right from one strip to the another we have,
\begin{align}
    H_{AB(BA)}^{LS}=H_{AB(BA)}^{RS}=\begin{bmatrix}
        H_{AB(BA)}^{PMSC} &&&&&\\
        &H_{AB(BA)}^{PMSC}&&&&\\
        &&H_{AB(BA)}^{PMSC}&&&\\
        &&&..&&\\
        &&&&..&\\
        &&&&&H_{AB(BA)}^{PMSC}
    \end{bmatrix}.
\end{align}

Now we define $H^{AM}_{r,r'}$ for AM region. We will use subscript $0$, $x$, $y$, $xy$, $x\bar{y}$ for onsite energy, $x$ direction, $y$ direction nearest neighbor hopping, next nearest hopping in $x+y$ and $x-y$ direction respectively. So the Hamiltonian matrices are,
\begin{align}
    H^{AM}_{0}=&-\mu \tau_{z}s_{0}\\
    H^{AM}_{x}=&-t^{AM}\tau_{z}s_{0} - t_{j}^{AM} \cos(2\alpha)\tau_{z}s_{z} - \frac{\lambda }{2ia}(\sin{\alpha}\tau_{0}s_{x}+\cos{\alpha}\tau_{z}s_{y})\\
    H^{AM}_{y}=&-t^{AM}\tau_{z}s_{0} + t_{j}^{AM} \cos(2\alpha)\tau_{z}s_{z} - \frac{\lambda }{2ia}(\sin{\alpha}\tau_{z}s_{y} - \cos{\alpha}\tau_{0}s_{x})\\
    H_{xy}^{AM}=&-H_{x\bar{y}}^{AM}= -\frac{t^{AM}_{j}}{2}\sin(2\alpha)\tau_{z}s_{z}\\
\end{align}
where $\tau_i$ are pauli matrices acting on particle-hole space for $i\in 0,x,y,z$. Next, the on strip Hamiltonian matrix can be written as,
\begin{align}
    H_{11}^{AM}=\begin{bmatrix}
        H^{AM}_{0}&{H^{AM}_{y}}&&&&\\
        {H^{AM}_{y}}^{\dagger} & H^{AM}_{0}&{H^{AM}_{y}}& & &\\
        &  {H^{AM}_{y}}^{\dagger} & H^{AM}_{0}&{H^{AM}_{y}}& & \\
        &&&..&&\\
        &&&&..{H^{AM}_{y}}\\
        &&&{H^{AM}_{y}}^{\dagger} & H^{AM}_{0}
    \end{bmatrix}
\end{align}
and for hopping to the right strip in AM we have,
\begin{align}
     H_{12}^{AM}=\begin{bmatrix}
        H^{AM}_{x}&H^{AM}_{xy}&&&&\\
        {H^{AM}_{x\bar{y}}} & H^{AM}_{x}&H^{AM}_{xy}& & &\\
        &  {H^{AM}_{x\bar{y}}} & H^{AM}_{x}&H^{AM}_{xy}& & \\
        &&&..&&\\
        &&&&..H^{AM}_{xy}\\
        &&&{H^{AM}_{x\bar{y}}} & H^{AM}_{x}
    \end{bmatrix}
\end{align}.

The coupling of PMSC with AM, Eq. [\ref{eq10}],  at left junction and right junction is given as  $H^{C_{L(R)}}=t_c \tau_{z}s_{0}$ where $t_c$ is hopping amplitude. The strip hopping Hamiltonian matrix corresponding to the coupling is written as,  
\begin{align}
    \tilde{H}^{C_{L(R)}}=\begin{bmatrix}
        H^{C_{L(R)}}&&&&&\\
        &H^{C_{L(R)}}&&&&\\
        &&H^{C_{L(R)}}&&&\\
        &&&..&&\\
        &&&&..&\\
        &&&&&H^{C_{L(R)}}
    \end{bmatrix}.
\end{align} 

For the second configuration given in Fig.[\ref{fig5a}] coupling matrix $H^{C_{L(R)}}$ have different shape, given both sublattice in left SC is connected to single nearest lattice point in AM[\onlinecite{Fukaya2025}]. Where,
\begin{align}
    H^{C_{L(R)}}=t_{c}\tau_{z}\otimes s_{0}\otimes\begin{bmatrix}
        1\\1
    \end{bmatrix}.
    \label{A17}
\end{align}

\end{widetext}

\section{Algorithm}
\label{appB}
We now describe the algorithm used to obtain the CPR. Since the junction is oriented along the $x$-axis, the system can be viewed as a set of one-dimensional lattice strips extending along the $y$-direction, defined at each lattice point in $x$. Consequently, the junction can be treated as an effective one-dimensional system, where each site along $x$ represents an extended strip in the $y$-direction. Henceforth, a given $x$ coordinate refers to the entire strip along the $y$-direction. In other words, whenever we refer to a particular $x$ site, the corresponding full strip is implicitly included. The nonlocal retarded and advanced Green’s functions can then be expressed in terms of the local Green’s functions as follows:

\begin{align}
    G^{R(A)}_{LJ}(E)=&G^{R(A)}_{AM}(E)({\tilde{H}^{C_{L}})^{\dagger}}g^{R(A)}_{LSC}(E).     
\end{align}
Where $G_{AM}^{R(A)}(E)$ and $g_{LSC}^{R(A)}(E)$ are the retarded (advance) Green's function defined at the left junction surface of AM and left SC respectively. While $G_{AM}^{R}(E)$ is defined at $x=0$ which  is coupled to the whole junction, $g_{LSC}^{R}(E)$ on other hand is retarded(advance) surface Green's function at $x=-a$ of isolated left SC lead.  $G_{AM}^{R(A)}(E)$ and $g_{LSC}^{R(A)}(E)$ can in turn be find using recursive algorithm. For that we define $H_{11}^{AM}$ which is  Hamiltonian matrix representing ``on strip" hopping and on site potential,  $H_{12}^{AM}$ and $H_{21}^{AM}$ are Hamiltonian matrix for hopping from one strip to it's right and left strip respectively in AM region. Same terminology can be used for the PMSC region also but it has different sublattices along $x-$axis, we use, $H^{LS}_{AA}$  and $H^{LS}_{BB}$ for ``on-strip" Hamiltonian matrix for the strip of `A' and `B' sublattice. For hopping to the right strip we use $H^{LS}_{AB}$ and $H^{LS}_{BA}$  from `A' to `B' and `B' to `A' respectively.\\                      
First, we use forward procedure [\onlinecite{Do_2014}] to get $g_{LSC}^{R}(E)$, where we start with left most strip in left SC lead which is $H^{LS}_{AA}$ and calculate the retarded Green's function of that strip as,
\begin{align}
    g^{R}_{1}=&[\mathbf{1}E+i\zeta - H^{LS}_{AA}]^{-1}
\end{align}
where $\zeta$ is infinitesimal number to keep it from divergence and $\mathbf{1}$ is identity matrix of corresponding order. For second strip which is coupled to the first one its given as,
\begin{align}
    g^{R}_{2}=&[\mathbf{1}E+i\zeta - H^{LS}_{BB} - H_{AB}^{LS} g_{1}^{R}{H_{AB}^{LS}}^{\dagger}]^{-1}
\end{align}
and we keep going for every next strip up to the last one which is coupled to whole system to the left,
\begin{align}
    g^{R}_{LSC}=&[\mathbf{1}E+i\zeta - H^{LS}_{BB} - H_{AB}^{LS} g_{N_{x}^{L}-1}^{R}{H_{AB}^{LS}}^{\dagger}]^{-1}
\end{align}
Now we will move to use backward procedure [\onlinecite{Do_2014}] to get $G_{AM}^{R}(E)$. Similar to last procedure we consider right most strip of right PMSC. So, the green's function of that strip is given as,
\begin{align}
    G_{N_{x}^{R}+N_{x}^{AM}}^{R}(E)=& [\mathbf{1}E+i\zeta - H_{BB}^{RS}]^{-1}
\end{align}
for next strip to the left coupled to last one is,
\begin{align}
\nonumber
     G_{N_{x}^{R}+N_{x}^{AM}-1}^{R}(E)=& [\mathbf{1}E+i\zeta - H_{AA}^{RS} - \\&{H_{AB}^{RS}}^{\dagger} G_{N_{x}^{R}+N_{x}^{AM}}^{R}H_{AB}^{LS}]^{-1}
\end{align}
on keep going to the left and coupling the full system to its right we get the Green's function for $x=0$ strip which is given by,
\begin{align}
\nonumber
    G_{AM}^{R}(E)= &[\mathbf{1}E+i\zeta - H_{11}^{AM} - {H_{12}^{AM}}^{\dagger} G_{1}^{R}H_{12}^{AM}\\
    &-(\tilde{H}^{C_{L}})^{\dagger} g^{R}_{LSC}{\tilde{H}^{C_{L}}}]^{-1}
    \label{eq17}
\end{align}
We use $ g^{R}_{LSC}$ in Eq. [\ref{eq17}] to couple it to the left PMSC. Now we have all matrices required to compute current given in Eq. [\ref{5}]. 

\section{Symmetry Transformation Matrix}
\label{app C}
Here we define general symmetry transformation matrix including sublattice degree of freedom which can be removed for AM region. For TRS the matrix is given as [\onlinecite{PhysRevB.104.134514}] ,
\begin{align}
    \mathcal{T}=\begin{bmatrix}
        -i s_y\rho_{0} &0\\
        0 & -is_y\rho_{0}
    \end{bmatrix}\mathcal{K}
\end{align}
where s and $\rho$ apply on spin and sublattice space and $\mathcal{K}$ is complex conjugation operator. For mirror symmetry transformation in $xz$ plane,
\begin{align}
    M_{xz}=\begin{bmatrix}
        is_{y}\rho_{0} & 0 \\
        0  &  i s_y\rho_0
    \end{bmatrix}\mathcal{P_y}
\end{align}
where $\mathcal{P_y}$ act on real space, transform $y \rightarrow -y $ and $k_y\rightarrow-k_y$  For mirror symmetry transformation in $yz$ plane,
\begin{align}
    M_{yz}=\begin{bmatrix}
        is_x\rho_{x} & 0\\
        0 & -is_x\rho_x
    \end{bmatrix}\mathcal{P_{x}}
\end{align}
where $\mathcal{P_x}$ act on real space, transform $x\rightarrow -x$ and $k_x\rightarrow-k_x$ For $\pi$ spin rotation about $y-$axis,
\begin{align}
    R_{y}=\begin{bmatrix}
        -is_y\rho_0 & 0\\
        0 & -is_y\rho_0
    \end{bmatrix}
\end{align}

For $\pi$ spin rotation about $z-$axis,
\begin{align}
    R_{z}=\begin{bmatrix}
        -is_z\rho_{0} & 0\\
        0 & is_z\rho_0
    \end{bmatrix}.
\end{align}
\bibliography{biblo}

\end{document}